\documentclass[prd,twocolumn,preprintnumbers]{revtex4}

\usepackage{amsmath}
\usepackage{epsfig}
\usepackage{graphicx}
\usepackage{color}
\usepackage[normalem]{ulem}
 \usepackage{url}
\usepackage[breaklinks, plainpages=false, colorlinks=true, anchorcolor=cyan, linkcolor=red, citecolor=cyan, urlcolor=magenta, bookmarks=false]{hyperref}

\usepackage[caption=false]{subfig}

\begin{document}
\renewcommand{\thefigure}{\arabic{figure}}
\setcounter{figure}{0}

 \def\I{{\rm i}}
 \def\E{{\rm e}}
 \def\D{{\rm d}}

\bibliographystyle{apsrev}

\title{Bayesian power spectral estimation of gravitational wave detector noise revisited}

\author{Toral Gupta}
\affiliation{eXtreme Gravity Institute, Department of Physics, Montana State University, Bozeman, Montana 59717, USA}

\author{Neil J. Cornish}
\affiliation{eXtreme Gravity Institute, Department of Physics, Montana State University, Bozeman, Montana 59717, USA}

\begin{abstract} 
The analysis of gravitational wave interferometer data requires estimates for the noise covariance matrix. For stationary noise, this amounts to estimating the power spectrum. Classical methods such as Welch averaging are used in many analyses, but this method require large stretches of ``off-source'' data, where the assumption of stationarity may break down. For this reason, Bayesian spectral estimates using only ``on-source'' data are becoming more widely used, but the Bayesian approach tends to be slower, and more computationally expensive than classical methods. Here we introduce numerous improvements in speed and performance for the {\tt BayesWave} trans-dimensional  Bayesian spectral estimation algorithm, and introduce a new, low-latency fixed dimension Bayesian spectral estimation algorithm, {\tt FastSpec}, which serves as both a starting point for the {\tt BayesWave} analysis, and as a stand-alone fast spectral estimation tool. The performance of the Welch, {\tt BayesWave} and  {\tt FastSpec} algorithms are compared by applying statistical tests for normality to the whitened frequency domain data. Bayesian spectral estimation methods are shown to significantly outperform the classical approach.
\end{abstract}

\maketitle

\section{Introduction}

As the sensitivity of ground based gravitational wave detectors improves, the number of events increases, as does the time span over which the signals are detectable. To be able to handle the ever increasing event rate and signal duration, fast analysis methods are needed. A first step in any analysis is an estimate of the noise properties, which for Gaussian noise is given by the noise covariance matrix. For stationary, Gaussian noise the noise properties are summarized by the power spectral density (PSD). In this work, we describe a new, fast Bayesian spectral estimation algorithm - {\tt FastSpec} that can be used for low latency analyses, and we also introduce numerous improvements and speed-ups to the higher fidelity, but slower, {\tt BayesWave} trans-dimensional spectral estimation algorithm.

In reality, the data from the LIGO~\cite{LIGOScientific:2014pky,aLIGO:2020wna} and Virgo~\cite{VIRGO:2014yos,Virgo:2019juy} detectors are neither perfectly Gaussian or stationary~\cite{LIGOScientific:2019hgc}, and more complex models are needed to handle noise transients (glitches)~\cite{Cornish:2014kda,Cornish:2020dwh,Chatziioannou:2021ezd,Hourihane:2022doe,LIGO:2021ppb} and variations in the noise level with time~\cite{Cornish:2020odn,Digman:2022igm}. However, spectral estimates still play an important role in these analyses. For example, in applications that use a wavelet domain analysis to handle non-stationary noise~\cite{Cornish:2020odn}, the usual spectral estimates describe the average noise levels as a function of frequency, and provide a good starting for estimating the dynamic power spectrum, which varies with both frequency and time. 

Currently, most low latency analyses of LIGO/Virgo/Kagra (LVK) data, including online searches for signals from binary merger events~\cite{Nitz:2018rgo,DalCanton:2020vpm,Ewing:2023qqe,Aubin:2020goo,PhysRevD.105.024023} and sky localization for electromagnetic follow-up~\cite{Singer:2015ema}, use classical spectral estimation techniques such as Welch averaging to estimate the noise power spectrum. For example, the PyCBC Live algorithm~\cite{Nitz:2018rgo,DalCanton:2020vpm} uses Welch's method~\cite{1161901} and median averaging to estimate the power spectrum. Typical settings for the Welch spectral estimates use 32 segments of ``off-source'' data, with same duration as the on-source window, from times immediately preceding the analysis segment. Drifts in the noise levels and impulsive noise transients can bias these off-source estimates~\cite{Chatziioannou:2019zvs}. 
Moreover, even for perfectly stationary Gaussian noise, it is known that data whitened using Welch's method does not follow a standard zero mean, unit variance ${\cal N}(0,1)$ normal distribution~\cite{Rover:2008yp,Talbot:2020auc}. Welch's method with mean averaging results in a Student's-$t$ distribution, while median averaging results in another more complicated distribution with even heavier tails~\cite{Talbot:2020auc}. Both of these distributions approach a ${\cal N}(0,1)$ normal distribution in the limit that the number of data segments being averaged over is large, but using more segments makes the method more susceptible to non-stationarity. Using Welch's method to estimate the power spectral has been shown to lead to biased parameter estimation for compact binary mergers~\cite{Chatziioannou:2019zvs,Talbot:2020auc}.

For these reasons, ``on-source'' Bayesian spectral estimates, such as those provided by the {\tt BayesWave} algorithm~\cite{Cornish:2014kda,Cornish:2020dwh,Littenberg:2014oda}, have been adopted as the preferred method for the more in-depth, longer latency parameter estimation studies that are performed on candidate signals from the search pipelines (see e.g.~\cite{LIGOScientific:2021usb,KAGRA:2021vkt}). The Bayesian spectral estimates have been shown to yield unbiased and statistically consistent results, with whitened residuals that follow a ${\cal N}(0,1)$ normal distribution~\cite{Chatziioannou:2019zvs}. 

Ideally, the spectral estimation should be performed concurrently with the source inference. This procedure was introduced in the original {\tt BayesWave} analysis of minimally modelled sources using a wavelet based signal~\cite{Cornish:2014kda}. The {\tt BayesWave} analysis has now been extended to include template based models for compact binary mergers, with full marginalization over the spectral model~\cite{Hourihane:2022doe}. The joint modeling of the noise spectrum and the signal was found to have a relatively small impact on parameter estimation for compact binary mergers~\cite{Plunkett:2022zmx}, which supported the findings of an earlier study that took multiple fair draws from the {\tt BayesWave} spectral model, and marginalized over these by alternating which one was used when computing the likelihood~\cite{Biscoveanu:2020kat} (using off-source Welch spectral estimates rather than the on-source {\tt BayesWave} model was shown to cause a much larger bias in the parameter estimation~\cite{Chatziioannou:2019zvs}). While sequential and joint spectral modeling was found to produce similar parameter recovery at current sensitivity levels, the differences will become more significant as the detector sensitivities improve. It is also worth noting that joint signal and noise modeling will be essential for next generation detectors on the ground and in space, as there will no signal-free off-source data to work with, and the signals will be a significant contribution to the total power. Joint signal and noise modeling will be a key element of the {\em Global Fit} analysis~\cite{PhysRevD.72.043005,Cornish:2005qw,Littenberg:2020bxy,Littenberg:2023xpl} for the future Laser Interferometer Space Antenna (LISA).

While Bayesian spectral estimates are more reliable, they come with a higher computational cost. For example, producing a spectral estimate for 128 seconds of LIGO data covering frequencies up to 4096 Hz, which are standard choices for binary Neutron star merger events, would take a day or so with the version of {\tt BayesWave} used during the LVK third observing run (O3). In preparation for the fourth observing run (O4), a new low-latency spectral estimation algorithm, {\tt FastSpec}, has been developed which can be used to provide standalone low latency Bayesian spectral estimates, and also provides a starting point for the full {\tt BayesWave} analysis. In addition, numerous changes were made to the {\tt BayesWave} algorithm that improved the quality of the spectral estimates and reduced the computational cost. The combination of the better starting solution from {\tt FastSpec}, faster likelihood evaluations and improved sampling, result in spectral estimates that take one to two orders of magnitude less time using the O4 version of {\tt BayesWave} than with the O3 version. The {\tt FastSpec} analysis is faster still, with run times that are roughly equal to the duration of the data segments being analyzed.

The paper is organized as follows: Section~\ref{specest} reviews the classical and Bayesian spectral estimation methods used in LVK data analysis. Section~\ref{fastspec} describes the {\tt FastSpec} low-latency spectral estimation algorithm. Section~\ref{bw} describes the improvements made to the {\tt BayesWave} trans-dimensional Bayesian spectral estimation algorithm. Section~\ref{results} compares the performance of Welch's method to the {\tt FastSpec} and {\tt BayesWave} algorithms using data from the LVK third observing run~\cite{KAGRA:2023pio}. Section~\ref{discuss} provides a summary and a discussion of possible future improvements.

\section{Spectral Estimation}\label{specest}

A variety of classical and Bayesian techniques have been used to produce power spectral density estimates of the instrument noise for use in gravitational wave data analysis. Classical approaches include the aforementioned Welch's method~\cite{1161901}, and the fitting of Auto Regressive Moving Average models ${\rm ARMA}(p,q)$ with $p$ autoregressive terms and $q$ moving-average terms~\cite{ElenaCuoco_2001}. A variety of Bayesian spectral estimation methods have also been tried, including both non-parametric~\cite{Edwards_2018,Maturana_Russel_2021,10.1214/18-BA1126} and parametric models~\cite{Littenberg:2014oda,Chatziioannou:2019zvs}. A more extensive review of the various approaches can be found in Ref.~\cite{RevModPhys.94.025001}.
The focus here will be on the two main methods that are currently used in LVK data analysis, Welch's method and Bayesian parametric modeling using {\tt BayesWave}.

Welch's method uses multiple, often overlapping, time segments. The data in each segment are windowed then Fourier transformed to produce a collection of periodograms. By taking either the mean or median of the periodograms at each frequency yields an estimate for the power spectrum. Taking the median is more robust against outliers, such as noise transients that might impact some of the data segments. Welch's method makes no assumptions about the functional shape of the spectrum, but it does rely on the data being stationary. If the data are perfectly stationary, the Welch estimate improves with the number of segments used, and the distribution of the whitened residuals approaches at ${\cal N}(0,1)$ distribution~\cite{Rover:2008yp,Talbot:2020auc}. In practice, the LVK data are non-stationary over timescales of minutes or less, so for reasons of both computational efficiency and to avoid biases from non-stationarity, a relatively small number of segments are used. The Welch estimates are referred to as ``off-source'' estimates since the data segments are usually taken from times immediately before the analysis segment. 

\begin{figure}[htp]
\includegraphics[width=0.48\textwidth]{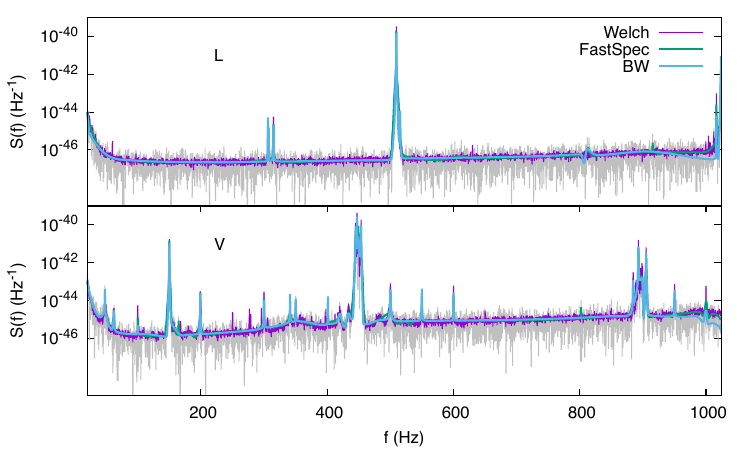} 
\caption{\label{fig:psds} Comparison of PSD estimates for 4 seconds of O3 LIGO Livingston and Virgo data ending at UTC 0:0:0, January 1, 2020 (GPS time 1261872018). The {\tt FastSpec} and {\tt BayesWave} estimates are almost identical, and are in good agreement with the Welch estimate. Periodograms of the data used in the {\tt FastSpec} analysis are shown in grey. Note that the apparent discrepancy between the {\tt BayesWave} and {\tt FastSpec} analyses above 950 Hz are due to differences in the bandpass filters used in each analysis.}
\end{figure}

The {\tt BayesWave} algorithm models the power spectral density function $S(f)$ using a combination of splines for the broadband features, and Lorentzians for the spectral lines. Crucially, the number of spline knots and the number of Lorentizian line features are dynamically determined using a trans-dimensional Reversible Jump Markov Chain Monte Carlo (RJMCMC) routine. The analysis uses the Whittle likelihood, the log of which can be written as
\begin{equation}
\log p ({\bf n} | S(f)) = -\sum_j \left[ \frac{|\tilde{n}(f_j)|^2}{T S(f_j)} + \log\left(\pi T S(f_j)\right)\right]
\end{equation}
where $T$ is the observation time and $f_j = j/T$ are the discrete frequency bins. A detailed description of the {\tt BayesWave} spectral estimation method and the improvements that have been made are provided in Section~\ref{bw}. Another key feature of the {\tt BayesWave} algorithm is that, concurrent with the spectral estimation, any non-Gaussian excess caused by either noise transients or loud signals are subtracted from the data. The non-Gaussian features are modeled as a sum of Morlet-Gabor wavelets, where again, the number of wavelets is dynamically determined by a trans-dimensional RJMCMC sampler~\cite{Cornish:2014kda,Cornish:2020dwh}. The {\tt FastSpec} algorithm is designed to mimic the main elements of the {\tt BayesWave} algorithm while significantly reducing the computational cost by replacing fully trans-dimensional Bayesian inference with a mixture of deterministic data processing steps followed by a fixed dimension Bayesian refinement. A detailed description of the  {\tt FastSpec} algorithm can be found in Section~\ref{fastspec}.

Figure~\ref{fig:psds} compares PSD estimates from the Welch method, {\tt FastSpec} and {\tt BayesWave} applied to 4 seconds of O3 LIGO Livingston and Virgo data ending at UTC 0:0:0, January 1, 2020 (GPS time 1261872018). This data, and all the other data used in our analyses, was sourced from the LVK public data release~\cite{KAGRA:2023pio}, which uses the channels {\tt H/L1:DCS-CALIB\_STRAIN\_CLEAN-SUB60HZ\_C01} and {\tt V1:Hrec\_hoft\_16384Hz}.

All three methods produced similar estimates for this example. The Welch estimate is not very smooth since just 32 4-second segments of data from times immediately preceding the on-source segment were used in the averaging. Increasing the number of segments used in the averaging yields smoother estimates, but this also increases the computational cost and the impact of any non-stationarity in the data.

\section{Fast Bayesian Spectral Estimation}\label{fastspec}

The {\tt FastSpec} algorithm shares many of the attributes of the {\tt BayesWave} algorithm, but with a much reduced computational cost. Both perform ``on-source'' PSD estimation with subtraction of non-Gaussian features (either gravitational wave signals or instrumental noise transients). The main differences are that {\tt FastSpec} uses wavelet de-noising~\cite{1998BAMS...79...61T} to remove transients during an initial cleaning phase, while {\tt BayesWave} jointly infers the structure of the transients and the PSD model. Additionally, {\tt FastSpec} uses a fixed dimension MCMC sampler to refine the spline and line model, while {\tt BayesWave}  uses a trans-dimensional RJMCMC sampler.

\subsection{Initial Spectral Estimation and Wavelet De-noising}

\begin{figure}[htp]
\includegraphics[width=0.5\textwidth]{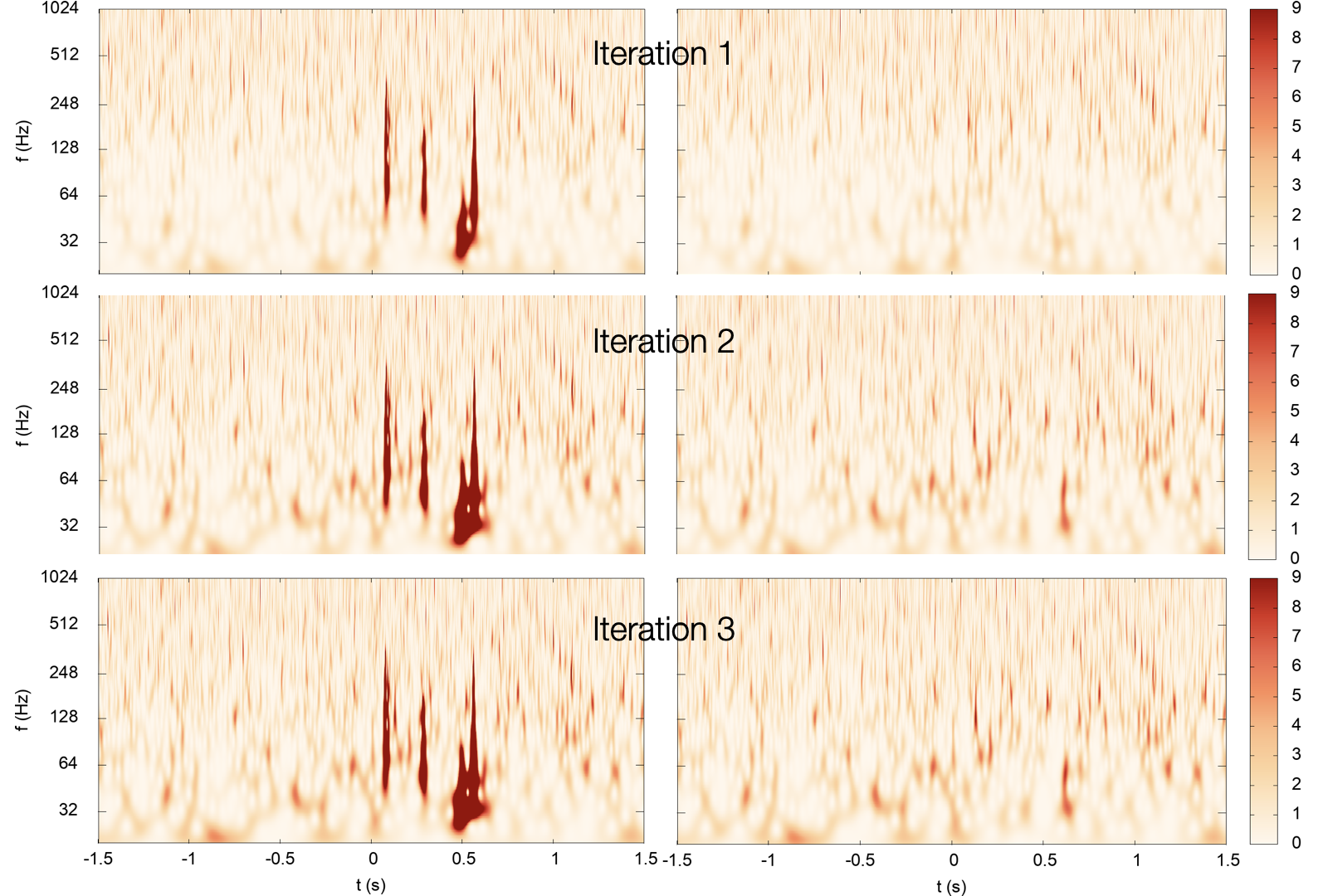} 
\caption{\label{fig:wdn} PSD estimation and Glitch reconstruction for the LIGO Hanford detector centered on GPS time 1263011016. The spectrograms on the left show successive stages in the initial data whitening, while the panels on the right show the corresponding de-noised data that is used in subsequent stages of the PSD estimation. Initially the three loud glitches distort the PSD estimation for frequencies below 300 Hz. Wavelet denoising is used to reconstruct and remove these glitches, resulting in improved PSD estimates that reveal the underlying Gaussian noise contribution. The ${\rm S/N}$ of the extracted non-Gaussian features at each iteration were $21.2$, $31.2$ and $32.1$ respectively. }
\end{figure}

The initial cleaning phase in {\tt FastSpec} is iterative and deterministic. The first step is to estimate the PSD using a running median of the periodogram. The width of the smoothing window is chosen to be wider than that of the line features, but smaller than scale over which the broadband noise varies appreciably. Window widths between 4 Hz and 16 Hz wide work well. Lines are identified as regions where the periodogram is a factor of 10 higher than the running median. The PSD model is then given by the value of the running median for regions away from the lines, and set equal to the periodogram value for regions in a line. This initial PSD estimate is then used to whiten the data. The data is then transformed to the time-frequency domain using continuous Morlet-Gabor (sine-Gaussian) wavelets (known as a Q-transform in the LVK literature). The wavelet quality factor, which fixes the width of the Gaussian envelope, is usually set at $Q=8$ since this value does a good job of capturing short duration transients. Wavelet de-noising~\cite{1998BAMS...79...61T} is then used to remove any features that exceed 3-sigma in the wavelet transform. If the signal-to-noise of the reconstructed non-Gaussian transient exceeds some threshold, (usually set at ${\rm S/N}=5$), the median plus line PSD estimation procedure is repeated on the cleaned data. This process is repeated until the ${\rm S/N}$ of the transient stabilizes, which is taken to be when the change in ${\rm S/N}$ between iterations is less than 5. The PSD model and the model for the non-Gaussian excess typically converge to a stable solution after a few iterations.

\begin{figure}[htp]
\includegraphics[width=0.48\textwidth]{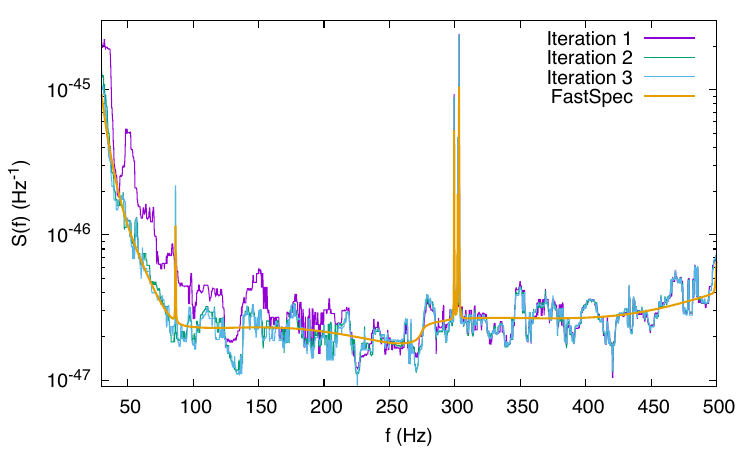} 
\caption{\label{fig:psd} Successive stages in the refinement of PSD estimate for 4 seconds of LIGO Hanford detector centered on GPS time 1263011016. The loud glitches bias the initial PSD estimate at frequencies below 300 Hz. The PSD estimate converges to a lower value as the wavelet de-noising procedure reconstructs and removes the glitches. The coarse initial PSD estimates based on a running median are compared to the final much smoother spectrum inferred by the subsequent Bayesian spectral modeling stage of the FastSpec algorithm.}
\end{figure}

Figure~\ref{fig:wdn} shows spectrograms of the data during the iterative wavelet de-noising procedure. Notice that the contrast in the Gaussian noise component for frequencies below 300 Hz grows as the fit for the noise transient is improved. Figure~\ref{fig:psd} shows how the running median plus line model for the PSD evolves during the iterative refinement of the glitch model.

\begin{figure}[htp]
\includegraphics[width=0.48\textwidth]{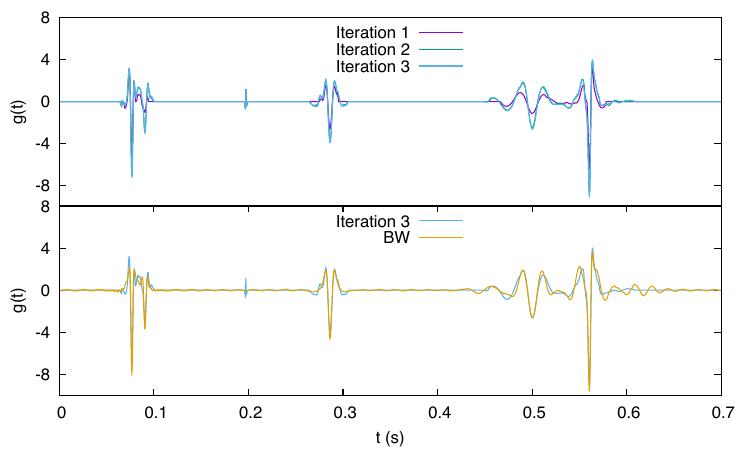} 
\caption{\label{fig:glitch} Glitch reconstruction for the LIGO Hanford detector centered on GPS time 1263011016. The top panel shows the refinement of glitch reconstruction during the iterative wavelet denoising procedure. The lower panel compares the final wavelet denoising glitch reconstruction to the BayesWave glitch reconstruction.}
\end{figure}

The plots in Figures~\ref{fig:wdn}, \ref{fig:psd} are for 4 seconds of LIGO Hanford data. The performance of the cleaning stage improves for longer data segments as there are more frequency samples in the window used to compute the running median, and isolated glitches or signals have a smaller impact on the initial PSD estimate, so the algorithm converges more rapidly. It is interesting to compare the glitch reconstruction from the {\tt FastSpec} cleaning phase (which takes a fraction of a second) with the median glitch reconstruction from the full {\tt BayesWave} analysis (which takes minutes). The upper panel of Figure~\ref{fig:glitch} shows the iterative refinement of the (whitened) glitch model in the time domain during the {\tt FastSpec} cleaning phase. In this particular case, the solution converged in three iterations. For louder glitches, such as the one that impacted the LIGO Livingston detector data for the BNS event GW170817~\cite{LIGOScientific:2018hze}, it can take four or five iterations to converge. The lower panel of Figure~\ref{fig:glitch} compares the final {\tt FastSpec} glitch reconstruction to the median {\tt BayesWave} reconstruction. The agreement is remarkable, especially considering that the two algorithms are very different.

\subsection{Bayesian Refinement}

The initial PSD estimates from the {\tt FastSpec} cleaning phase are not very smooth, so the next step is to refine the modeling using a fixed dimension version of the {\tt BayesWave} spline and line model.  Both models fit to the log of the PSD. The analysis is initialized by judiciously placing the spline knots and fitting the line features by Lorentzians. The spline knots are placed by computing linear least squares fits to the running-median portion of the initial log PSD estimate. Starting at the minimum frequency used in the analysis, frequency windows of increasing size are tried until the chi-squared of the linear fit exceeds a pre-set threshold. A spline knot is then placed at the end of the window, with amplitude set equal to the the running-median of the log PSD at that location. The same procedure is then repeated until the maximum frequency used in the analysis is reached. This approach puts a higher density of spline knots in regions where the spectrum is changing most rapidly. The next step is to fit Lorentzians to each line features in the log PSD from the cleaning stage. The location in frequency, width in frequency, and the the height of lines are used to set the initial parameters for the Lorentzians. With the model initialized, a fixed dimension MCMC is used to refine the spline and line models. The parameters are the locations of the spline knots, the height of the spline knots, and the location, width and height of the Lorentzians.  We found that the method used to initialize the spline and line models tended to put in more knots and lines than the transdimensional {\tt BayesWave}  model, so we introduced a ``poor man's'' version of the transdimensional step by periodically evaluating whether deleting a given spline knot or line changed the log likelihood by $\Delta {\rm log} L < -2$, and if it did not, that knot or line was removed. This pruning procedure is only applied during the burn-in phase since it violates detailed balance. After burn-in, the dimension of the spline and line models is fixed, and the analysis is fully Bayesian.

\section{Improvements to the BayesWave Algorithm}\label{bw}

The {\tt BayesWave} spectral estimation algorithm~\cite{Littenberg:2014oda} is used by the LIGO/Virgo/Kagra collaborations to produce point estimates for PSD that are used in parameter estimation studies of compact binary mergers. The on-source BayesWave PSD has been shown to allow for more accurate source parameter recovery than traditional off-source PSDs, such as those founding using Welch averaging~\cite{Chatziioannou:2019zvs}. During the break between the LVK third and fourth observing runs we took the opportunity to improve the BayesWave PSD estimation in several ways, resulting in an updated algorithm that is both faster and more effective at whitening the data.

The most significant change was incorporating the {\tt FastSpec} algorithm as a rapid pre-processing step that can provide a very good starting solution for the spline and line models.. This improves the convergence of both the {\tt BayesWave} glitch model and the PSD model. The {\tt BayesWave} glitch model benefits from having both a good starting solution for the wavelets~\cite{Cornish:2020dwh}, and by starting with {\tt FastSpec} PSDs that are robust against noise transients, making the separation between glitches and the Gaussian component of the noise more obvious. The {\tt FastSpec} PSD is now used to define the prior bounds on the PSD model by defining an envelope around the median PSD estimate. The envelope is chosen to be wide enough so that the {\tt BayesWave} PSD 99\% credible interval is well inside the envelope, yet narrow enough that the analysis converges quickly.

\begin{figure}[htp]
\includegraphics[width=0.48\textwidth]{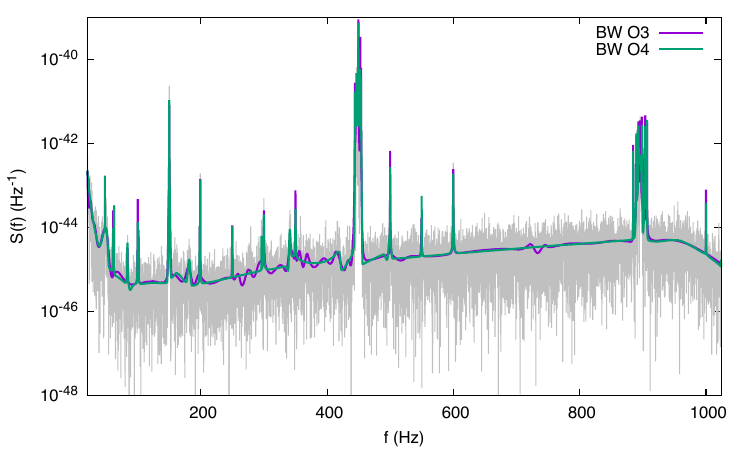} 
\caption{\label{fig:BWcomp} PSD estimates for 8 seconds of Virgo data centered on GPS time 1263011014. A periodogram of the data is shown in light gray, with spectral estimates from the version of BayesWave used in LVK observation run 3 (O3) and the version used in observation run 4 (O4). The O4 version uses Akima splines, and shows less variation than the O3 version, which uses Cubic splines.}
\end{figure}

Numerous changes were also made to the {\tt BayesWave} PSD model and the sampling procedure that improved the fidelity of the model and the convergence of the sampler. The original spline model used cubic splines and placed the spline control points (also called knots) on a discrete grid in frequency. The discrete frequency grid was used since cubic splines have a tendency to produce interpolating functions that suffer from large oscillations when the knots are spaced too closely together. In the original version, the location of the knots could only be changed in the trans-dimensional moves where knots were added or deleted.  In the updated version, cubic splines have been replaced by Akima splines. Akima splines produce a continuously differentiable sub-spline interpolation that is built from piecewise third order polynomials. Only data from the next neighbor points is used to determine the coefficients of the interpolation polynomial. Akima splines offer numerous advantages over cubic splines. Firstly, Akima are local, while cubic splines are computed using all the knots. Thus, when a single knot is updated the likelihood only has to be recomputed in the immediate vicinity of that knot, which reduces the computational cost. In addition, the locality means that Akima splines are not prone to develop unphysical oscillations, so we can dispense with the frequency grid and allow the knots to be placed at any position in frequency. Lateral moves are now allowed, so both the knot location and height can be varied in the fixed dimension MCMC updates. This added flexibility allows the spline knots to be optimally placed, which is especially important in regions where there are lines or other sharp features. Fig.~\ref{fig:BWcomp} compares the original O3 version {\tt BayesWave} PSDs to the updated O4 version for 8 seconds of Virgo data centered on GPS time 1263011014.  Virgo data was chosen to illustrate the difference between the O3 and O4 versions of {\tt BayesWave} because it is more challenging to fit due to the larger number of  spectral lines than the LIGO detectors. Note that the O4 version has far fewer wiggles than the O3 version thanks to the Akima splines. The O4 version allows for a prior to be set on the minimum spacing between points. When the minimum spacing is set to zero the prior on the number of spline knots is uniform. When a non-zero minimum spacing is used there are limitations on where the knots can be placed, resulting in an effective prior that exponentially decays. The results shown here use a minimum spacing of 2 Hz, though we found that setting the minimum to zero also produced good PSD models.

The LIGO and Virgo PSDs are known to have sharp line features from instrumental effects such as thermal excitation of the violin modes of the fibers that support the end mirrors. Many of the lines come from processes that can be model as damped-driven harmonic oscillators, which in the Fourier domain have power spectra that take the form of a Lorentzian. Each Lorentzian line is described by three parameters: central frequency $f_i$, quality factor $Q_i$ and height $A_i$, with the full model given by
\begin{equation}
S_L(f) = \sum_{i=0}^{N_L} \frac{A_i}{\left(\frac{f}{f_i}\right)^2 + Q_i^2 \left(1 - \left(\frac{f}{f_i}\right)^2 \right)^2 } \, .
\end{equation}
The number of lines, $N_L$, is explored by the trans-dimensional RJMCMC sampler. In the original version of {\tt BayesWave} the spectral model was written as the sum of the spline and line components $S(f) = S_S(f) + S_L(f)$, while in the new version the default is to write the model as the product $S(f) = S_S(f)\left(1 + S_L(f)\right)$. The advantage of the product model is that the line amplitudes are then always relative to the height of the spline model, making it easier to set priors on the line amplitudes. The switch from the sum to the product model has a much smaller impact on the resulting PSDs than the switch from cubic splines to Akima splines. The initial parameters for the line model come from the {\tt FastSpec} analysis. If the number of lines identified by {\tt FastSpec} exceeds the maximum number of lines specified for {\tt BayesWave}, the lines are sorted by amplitude and the highest amplitude lines are used.

Many changes were made to the sampler, including some bug fixes in the line and spline proposals that resulted in the O3 version of the code failing to recover the prior distribution on quantities such as the number of spline knots and the number of lines when the likelihood was set equal to a constant. These issues had been missed since the O3 version of the code used no-constant priors on the number of model components, which masked the issues with prior recovery. For example, the O3 version used an exponential prior on the number of lines, which hid the fact that the error in the proposal densities would otherwise lead to a runaway in the number of lines. In the O4 version we now use uniform priors on the number of knots and lines, and the updated sampler recovers these distributions when the likelihood is held fixed. 

The biggest changes to the sampling were related to the placement of the spline knots going from a discrete frequency grid to continuous. A new proposal was introduced for the trans-dimensional moves associated with adding or removing a spline knot. In addition to prior draws, a directed proposal was added that first randomly selects a frequency, then proposes that the amplitude of the spline knot is drawn from a Gaussian distribution centered on the current interpolated log PSD model at that frequency. This proposal involves no-trivial proposal densities for adding and deleting a knot. When deleting a knot, the interpolated log PSD model with the knot removed has to be calculated, with the log proposal density proportional to the squared difference in the log amplitude of the knot and the amplitude of the interpolated log PSD at the frequency of the knot that is being proposed for deletion. This targeted proposal has a higher acceptance rate than the simpler prior draws on the log amplitude. The updated sampler recovers the priors on all parameters when the likelihood is held constant.

\section{Performance Comparison}\label{results}

If the noise covariance matrix ${\bf C}$ has been correctly estimated from the data, then the data ${\bf d}$ can be whitened: ${\bf d}_w = {\bf W}^T {\bf d}$, where ${\bf W}$ is the Cholesky decompositoin of ${\bf C}^{-1} = {\bf W}{\bf W}^T$. For stationary data in the Fourier domain this procedure reduces to dividing the Fourier transform of the data $\tilde{\bf d}(f)$ by the amplitude spectral density $\sqrt{S(f)}$, such that $\tilde{\bf d}_w(f) = \tilde{\bf d}(f)/\sqrt{S(f)}$. Both the real and imaginary parts of $\tilde{\bf d}_w(f)$ should follow a ${\cal N}(0,1)$ normal distribution. This suggests a natural test for the PSD estimation methods - checking whether the whitened Fourier domain data is consistent with draws from a ${\cal N}(0,1)$ normal distribution. We chose to use two versions of the Anderson-Darling test, the first being the version that assumes the mean and variance are known, and the second being the version which makes no assumptions about the mean or variance, instead estimating these directly from the data samples. The latter ${\cal N}(\mu,\sigma)$ version is a stronger test of normality, while the ${\cal N}(0,1)$ test is useful in that it checks that the overall scaling is correct. We refer to the ${\cal N}(\mu,\sigma)$ version as the normality test.

\begin{figure}[htp]
\includegraphics[width=0.48\textwidth]{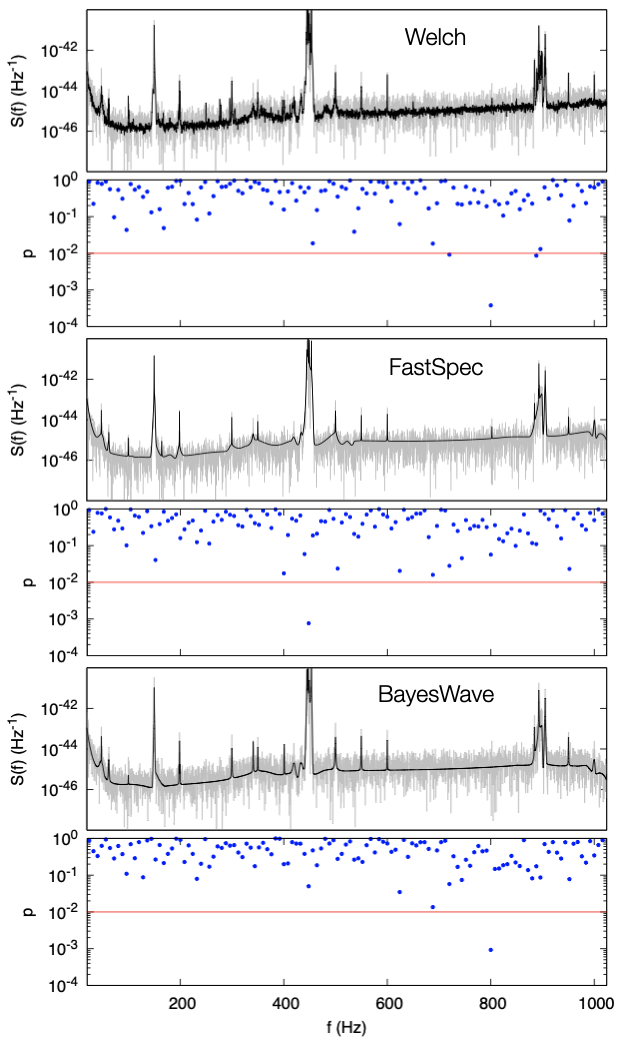} 
\caption{\label{fig:ADex} PSD estimates and Anderson-Darling normality test p-values applied to 4 seconds of Virgo data starting at GPS time 1261967014. The Anderson-Darling normality test was applied to data that was  whitened with the corresponding PSD estimates. The p-values are shown for bands of width 8 Hz, each containing $8 \times 4 \times 2 = 64$ samples. Applying the Anderson-Darling normality test to the full 1000 Hz band yielded p-values of:  Welch $p=0.40$, FastSpec $p = 0.79$, BayesWave $p = 0.92$. In this example, all three spectral estimation methods produced whitened data that are consistent with Gaussian noise.}
\end{figure}

Fig.~\ref{fig:ADex} compares the PSD estimates and Anderson-Darling normality test p-values applied to 4 seconds of Virgo data starting at GPS time 1261967014. Rather than just applying the normality test to the entire frequency range, we instead compute p-values for successive frequency bands of width 8 Hz. This helps to identify regions where the spectral estimates might be struggling. In the example shown in Fig.~\ref{fig:ADex}, all three spectral estimation methods produced whitened data that passes both version of the Anderson-Darling for the full bandwidth. The {\tt BayesWave} and Welch methods both fail the normality test in the 8 Hz window around 800 Hz, while the {\tt FastSpec} analysis fails the normality test in the 8 Hz region around 440 Hz. Note that the {\tt FastSpec} analysis picked up a small line at 800 Hz, while {\tt BayesWave} did not find the line to significant. The 440 Hz band where the {\tt FastSpec} analysis failed the normality test includes a very strong line.
In almost all examples we looked at, very low p-values were found to be associated with frequency bands containing spectral lines.

Note that failing the normality test does not necessarily imply that the PSD estimation procedure is incorrect as the data might not be Gaussian distributed. For example, if a loud signal or a loud glitch were present in the data we would not expect it to be Gaussian. Additionally, some instrument lines are know to drift (move around in frequency over time), which the models do not account for. The {\tt FastSpec} and {\tt BayesWave} analyses have an advantage since they remove non-Gaussian features when estimating the PSDs, and the normality tests are applied to the residuals from the analyses. As a result, the {\tt FastSpec} and {\tt BayesWave} analyses typically pass the normality tests even when very loud glitches are present in the original data.

\begin{figure}[htp]
\includegraphics[width=0.48\textwidth]{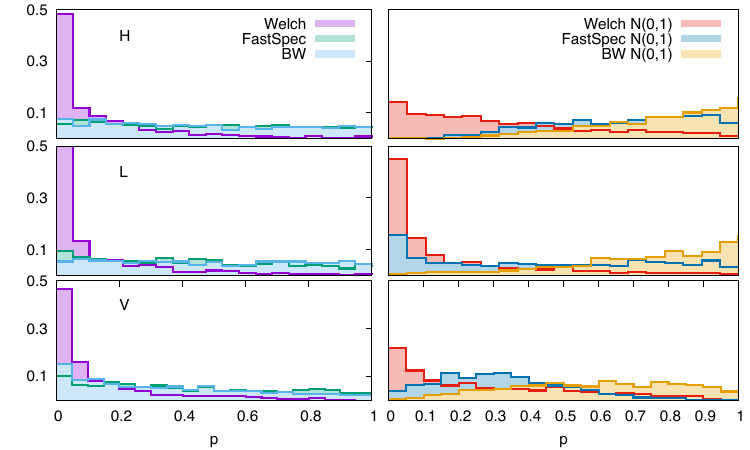} 
\caption{\label{fig:pvals4} The distribution of Anderson-Darling normality test p-values and Anderson-Darling ${\cal N}(0,1)$ test p-values for one thousand 4 second segments of triple co-incident LIGO Hanford, Livingston and Virgo data, starting from UTC 0:0:0, January 1, 2020 (GPS time 1261872018).  The data segments were separated by at least 1000 seconds. For each 4-second segment, the data were whitened using three methods, Welch, FastSpec and BayesWave, and p-values were computed for data between 20 Hz and 1020 Hz. The p-value distribution should be flat for data drawn from a Gaussian distribution. The Welch method failed both tests, whereas FastSpec and BayesWave, while still showing some bias, performed far better.}
\end{figure}

The performance of the various spectral estimation techniques was tested using data from the third observing run of the LIGO and Virgo detectors. The first set of tests used one thousand 4 second data segments starting at UTC 0:0:0, January 1, 2020. The data segments were spaced by 1000 seconds. The Welch averaging was performed using utilities from the {\tt Bilby}~\cite{Ashton:2018jfp} Python package, which uses the Welch median sub-routines from the {\tt GWpy}~\cite{gwpy} Python package. The median averaging was performed using 32 non-overlapping 4 second segments of data immediately prior to the analysis segment. The {\tt FastSpec} and {\tt BayesWave} analyses only used the 4 seconds of data in the analysis window. All three analyses used a sample rate of 2048 Hz. The Anderson-Darling ${\cal N}(0,1)$ test and ${\cal N}(\mu,\sigma)$ normality test were applied to the real and imaginary whitened Fourier domain data between 20 Hz and 1020 Hz. Histograms of the p-values for these tests are shown in Fig.~\ref{fig:pvals4}. The p-value distribution should be flat for data drawn from a Gaussian distribution. The Welch method failed both tests in all three detectors. {\tt FastSpec} and {\tt BayesWave} produced distributions that are consistent with an ${\cal N}(0,1)$ distribution for the LIGO Hanford and Livingston data, but showed a slight excess of low p-values for the Virgo data. The Virgo instrument tends to have many more spectral lines than the LIGO instruments, and when looking at the p-values in 8 Hz frequency windows, it was found that the low p-values corresponded to regions with strong spectral lines. The {\tt FastSpec} algorithm did not perform as well as {\tt BayesWave} under the  ${\cal N}(0,1)$ test, showing a slight excess of small p-values for both the LIGO Livingston and Virgo data. In contrast, the {\tt BayesWave} algorithm tended to produce p-values that were ``too good'', in other words, distributions that a skewed towards large p-values. Further investigation revealed that the bias in the p-values was due to using the median of the {\tt BayesWave} PSD distribution as a point estimate for the PSD. Repeating the test using fair draws from the posterior results in more uniform p-value distributions. This behavior was also noted in Ref.~\cite{Chatziioannou:2019zvs}. An in-depth investigation into the impact of using point estimates for PSD estimation can be found in the Appendix. 

\begin{figure}[htp]
\includegraphics[width=0.48\textwidth]{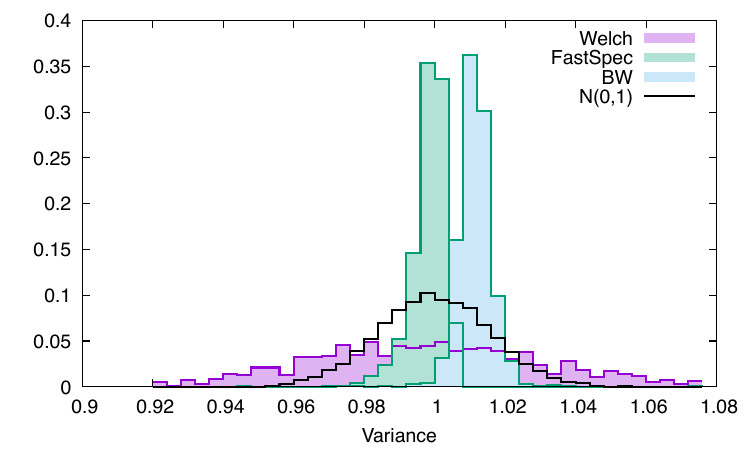} 
\caption{\label{fig:L1var} Histograms of the variances of the whiten LIGO Livingston data for the data segments shown in Figure~\ref{fig:pvals4}. The variance distribution for the same number of samples drawn from a ${\cal N}(0,1)$ Gaussian noise is shown for reference. The Welch method produces a spread that is wider than expected, while the FastSpec and BayesWave analyses produce variances that are narrower than expected, with slight biases either high or low.}
\end{figure}

Fig.~\ref{fig:L1var} focuses on the variance of the whitened data, which should be distributed around unity if the whitening is successful. The Welch method produced a distribution that is wider than expected and biased low, while the {\tt FastSpec} and {\tt BayesWave} algorithms produced much narrower distributions, with slight biases high or low. The narrowness of the distributions is due to the use of point estimates based on means or medians of the posterior distributions.

\begin{figure}[htp]
\includegraphics[width=0.48\textwidth]{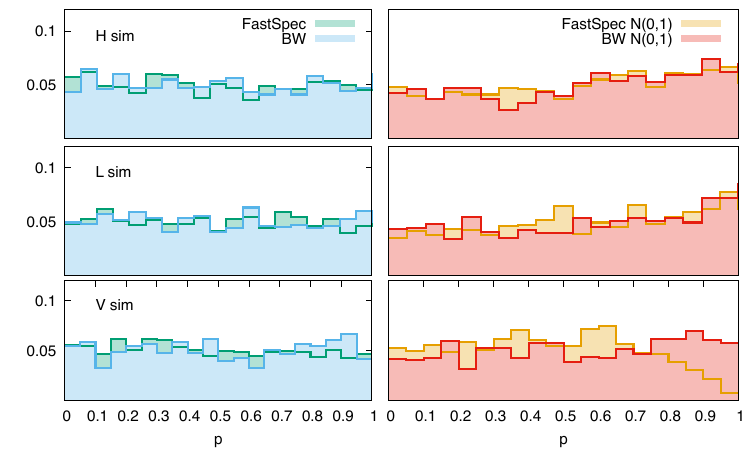} 
\caption{\label{fig:pvals4_fake} The distribution of Anderson-Darling normality test p-values and Anderson-Darling ${\cal N}(0,1)$ test p-values for one thousand 4 second segments of simulated data that were generated using the PSDs derived from the real data used in Figure~\ref{fig:pvals4}. Overall, both FastSpec and BayesWave passed this self-consistency check, with a slight bias toward high p-values for the ${\cal N}(0,1)$ test for reasons explained in the text.}
\end{figure}

When testing the algorithms on real data it is difficult to say if departures from normality are due to a problem with the whitening procedure, or intrinsic to the data itself. To test the internal self-consistency of the {\tt FastSpec} and {\tt BayesWave} algorithms we took the one-thousand PSD estimates from the analysis of the real data and generated one-thousand sets of simulated Gaussian noise drawn from those PSD estimates. The same analysis as before was then performed on the simulated data. Fig.~\ref{fig:pvals4_fake} shows the p-value distributions for the analyses performed on these simulated data sets. The whitened data passes the normality test in all detectors for both algorithms, while most of the p-value distributions for the ${\cal N}(0,1)$ test are biased towards high p-values for reasons explained in the Appendix. The exception is the {\tt FastSpec} analysis of the simulated Virgo data, where the large number of lines still presented a challenge.

\begin{figure}[htp]
\includegraphics[width=0.48\textwidth]{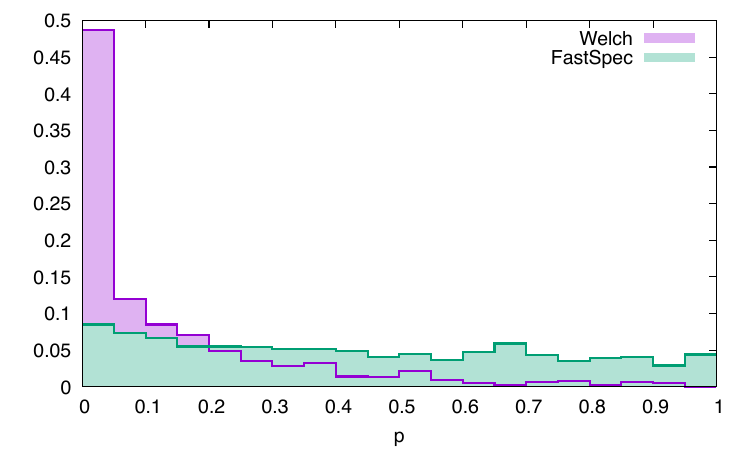} 
\caption{\label{fig:pvals16} The distribution of Anderson-Darling normality test p-values for one thousand 16 second segments of LIGO Hanford data, starting from GPS time 1261872018. The {\tt FastSpec} algorithm continues to significantly outperform the Welch method on these longer data segments. Similar results were found for longer segments of data, but non-stationarity starts to become an issue for minutes long segment lengths.}
\end{figure}

Returning to real data, a similar analysis was repeated using one thousand 16 second long segments of LIGO Hanford data, in this case focusing on the two low-latency analyses, Welch averaging and {\tt FastSpec}. The distribution of p-values for the ${\cal N}(0,1)$ Anderson-Darling test are shown in Fig.~\ref{fig:pvals16}. The {\tt FastSpec} algorithm performed very well on these longer data stretches, delivering a nearly uniform p-value distribution. The Welch method fared considerably worse, failing the ${\cal N}(0,1)$ test almost half the time. Similar tests were applied to longer segments of data, with the {\tt FastSpec} algorithm continuing to perform well, while the Welch method failed even more often. The main reason the Welch method fails is that LIGO and Virgo data are not stationary over extended periods of time. For example, the Welch method applied to 16 second segments used over 8 minutes of off-source data, and the noise levels fluctuate significantly on these timescales. 

\begin{figure}[htp]
\includegraphics[width=0.48\textwidth]{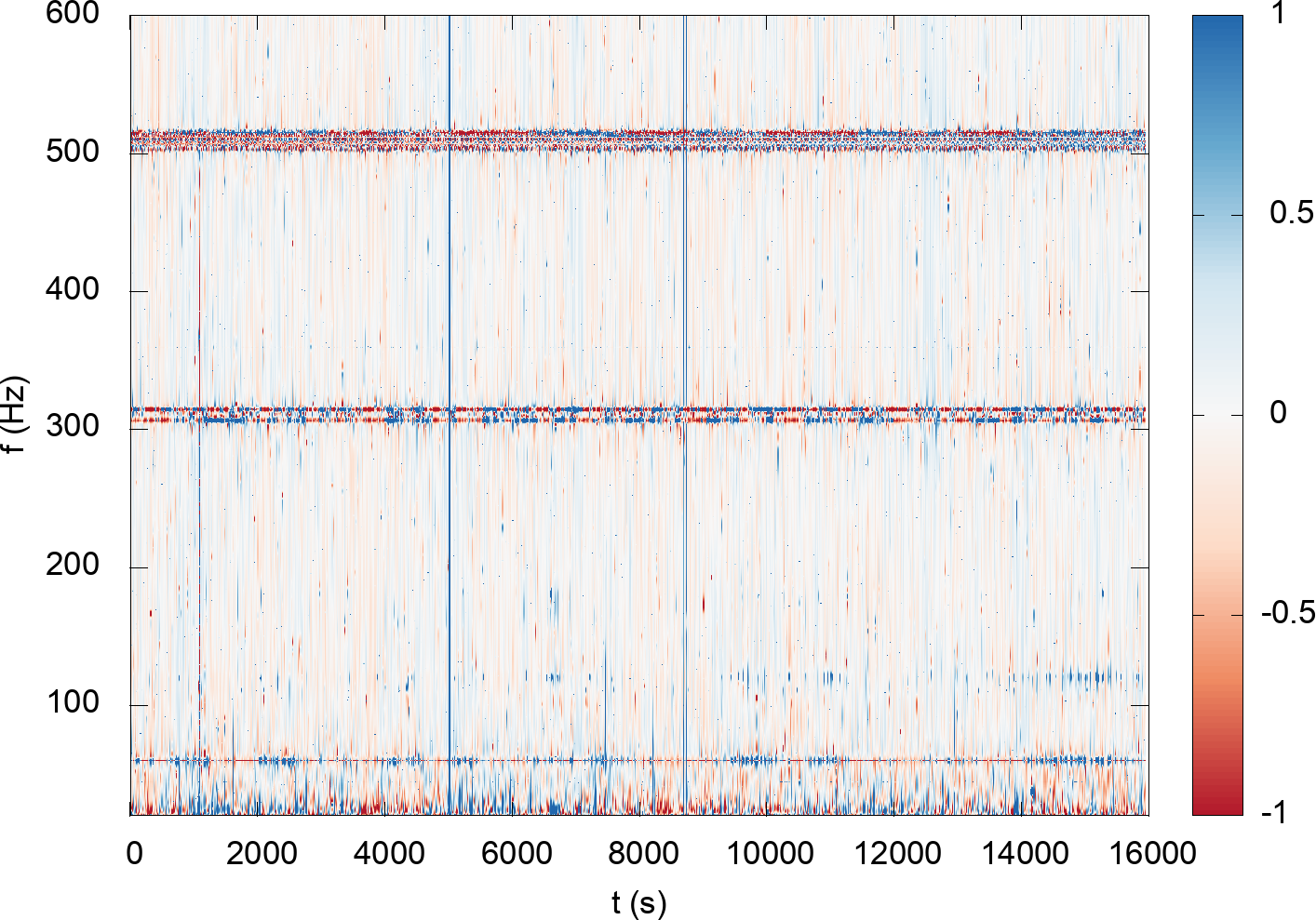} 
\caption{\label{fig:dspec} Non-stationarity in the PSD estimates for sequential 16 second stretches of LIGO Livingston data, starting at GPS time 1261872018. Here the PSD estimates for each segment were divided by the median of the PSD estimates for all 1000 segments, stacked vertically and plotted using a $\log_2$ color scale. Significant variation is seen around the dominant spectral lines (saturating the color scale), and for frequencies below 50 Hz. More moderate variation of order 10\%-50\% occurs across the spectrum.}
\end{figure}

The variation in the PSD over time is illustrated in Fig.~\ref{fig:dspec}. Here 1600 seconds of LIGO Livingston data, starting at GPS time 1261872018, were broken up into 16 second segments. {\tt FastSpec} was used to estimate the PSD for each segment. The data in each segment was then whitened using the median of the 1000 {\tt FastSpec} PSD estimates, stacked vertically and plotted using a $\log_2$ color scale to produce a spectrogram of the whitened data. Significant variation is seen around the strong spectral lines near 300 Hz and 500 Hz. The data below 30 Hz also shows significant variation. More moderate variation of order 10\%-50\% occurs across the spectrum. Several time slices also stand out as being especially noisy. It is these variations in the power spectrum that degrade the Welch estimates.

\section{Conclusions}~\label{discuss}

Bayesian parametric spectral estimation, especially when coupled with noise transient removal, is a powerful method for producing robust spectral estimates. Traditional methods such as Welch averaging use additional data outside the analysis window, which can cause problems when the noise levels fluctuate over time. Here we introduced a new low latency spectral estimation algorithm, {\tt FastSpec}, that while not quite as good as {\tt BayesWave}, should prove useful for low latency parameter estimation. We also made many changes to the original {\tt BayesWave} algorithm, improving the parametric model and significantly speeding up the analysis. The reduction in run time will be particularly important for long duration signals such as binary neutron star mergers, where the previous version took days to run while the new version takes hours - or just minutes if {\tt FastSpec} is used instead.

Going forward, both {\tt BayesWave} and {\tt FastSpec} could be improved by implementing better models for the spectral lines, but the gains would be marginal. Much more significant will be generalizing these models to handle non-stationary noise by switching from the frequency domain to the time-frequency domain. The standard LVK parameter estimation packages, such as {\tt Bilby}~\cite{Ashton:2018jfp}, use point estimates of the PSD. It would make sense to convert these analyses to being fully Bayesian, and marginalize over the PSD estimates jointly with the signal analysis, as is done in the {\tt BayesWave} signal analyses. Importing the fixed dimension sampler from {\tt FastSpec} should be relatively straightforward. The more adventurous could try incorporating the trans-dimensional {\tt BayesWave} model.

\section*{Acknowledgments}
The authors are grateful for the support provided by NSF award PHY 2207970. Many thanks to Tyson Littenberg for suggesting we try using Akima splines. Thanks also to Meg Millhouse, Sophie Hourihane and Katerina Chatziioannou for helpful discussions and tips on running {\tt BayesWave}. We also want to thank Carl-Johan Haster for acting as the LIGO code reviewer for this project. We appreciate Howard Desongs' help with integrating our changes into the production code. Many thanks to Eric Thrane, Carl-Johan Haster and Katerina Chatziioannou for providing feedback on an earlier draft of the manuscript. 

The authors are grateful for computational resources provided by the LIGO Laboratory and supported by National Science Foundation Grants PHY-0757058 and PHY-0823459. The data used in our study is based upon work supported by NSF's LIGO Laboratory which is a major facility fully funded by the National Science Foundation

\appendix*

\section{P-values for Point Estimates}

Ideally, PSD estimation and glitch fitting should be performed jointly with the parameter estimation for gravitational wave signals. This is what is done in the {\tt BayesWave} analyses for compact binary coalescences (CBCs)~\cite{Chatziioannou:2021ezd,Hourihane:2022doe}. However, other parameter estimation pipelines such as LALinference and Bilby do not yet have the facility to marginalize over the PSD estimates, instead relying on point estimates. Thus the Bayesian PSD estimation codes needs to supply some kind of point estimate from the full posterior distribution of PSD models. Possible choices include maximum likelihood estimates, maximum a posteriori estimates, a randomly chosen fair draw, or means or medians of the PSD estimates. For FastSpec we chose to use the average of the log PSD since it is easy to compute on the fly, and for BayesWave we choose to use the median log PSD since this estimate is more robust against outlier samples. Both the average and median estimates are examples of ``super resolution'' fits that do a better job of whitening the data than any random draw from posterior. The super resolution behavior is sometimes referred to as ``the wisdom of the crowd''. One downside of using super resolution estimates is that they do not produce uniform p-value distributions, instead they produce an over abundance of good fits, which skews the distributions towards higher p-values.

\begin{figure}[htp]
\includegraphics[width=0.48\textwidth]{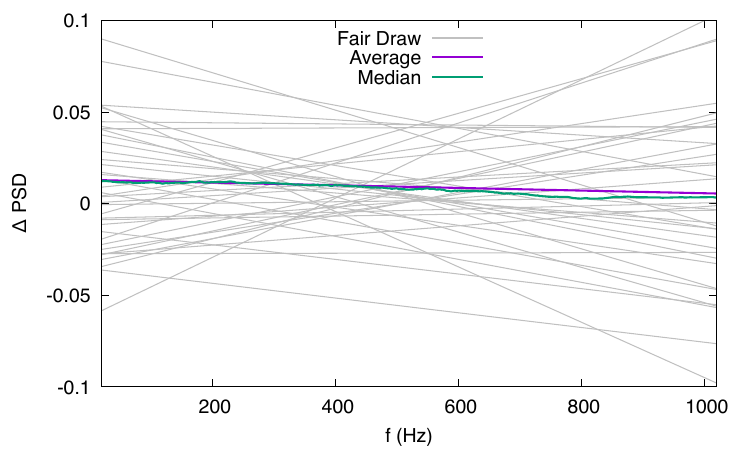} 
\caption{\label{fig:toy} Posterior draws with the mean and median post estimates for a toy model with a linearly increasing power spectrum. The true PSD used to generate the data has been subtracted to show the differences more clearly.}
\end{figure}

To illustrate this behavior using a simple model, we considered a toy version of Bayesian spectral estimation using simulated data with a linearly increasing power spectrum, and fixed dimension MCMC analysis with a linear model for the spectrum controlled by the two fixed points at the edges of the frequency range. Figure \ref{fig:toy} shows a collection of fair draws from the posterior samples, along with the mean and median of the posterior samples. The PSD model used to generate the data has been subtracted from these estimates to show the differences more clearly. Notice that both the mean and median track the simulated model far more accurately than a typical draw from the posterior. 

\begin{figure}[htp]
\includegraphics[width=0.48\textwidth]{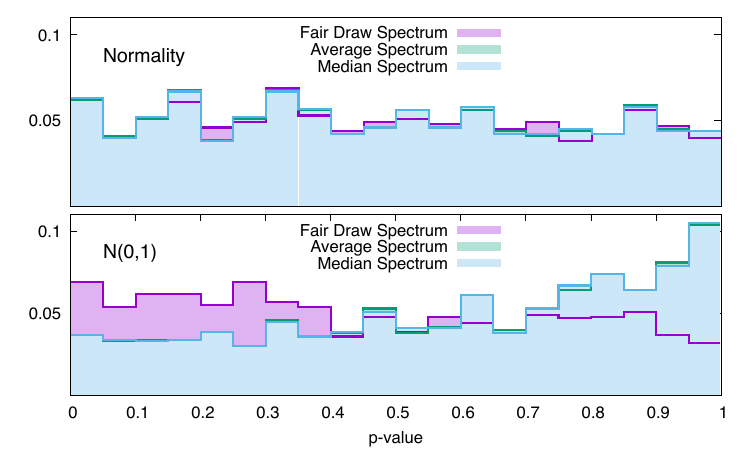} 
\caption{\label{fig:ptoy} Comparing the p-value distributions for the whitened toy model data using fair draw, mean and median PSD estimates.}
\end{figure}

Repeating the analysis on hundreds of simulated data sets and using the Anderson-Darling normality to produce p-values results on the distributions shown in Fig.~\ref{fig:ptoy}. When a fair draw from each posterior is used to whiten the data the resulting p-value distribution is uniform, while when the the mean and median of the posterior draws are used the p-value distribution is skewed towards larger values. 

\begin{figure}[htp]
\includegraphics[width=0.48\textwidth]{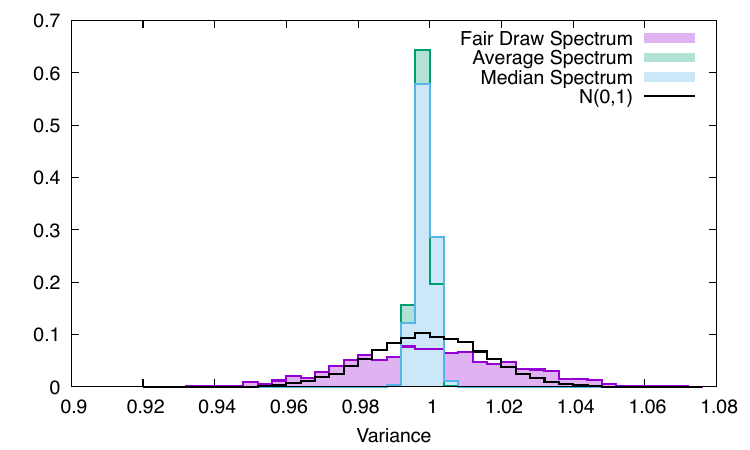} 
\caption{\label{fig:vtoy} Histograms of the variances of the whitened toy model data using fair draw, mean and median PSD estimates. The reference distribution for the same number of samples drawn from ${\cal N}(0,1)$ Gaussian noise is also shown. The fair draw PSDs yield distributions in line with the expected distribution, while the mean and median point estimates are narrowly peaked around unit variance.}
\end{figure}

The over performance of the super resolution fits using the mean and median also yield whitened data with a very narrow range of variances. Fig.~\ref{fig:vtoy} shows the variance of the data whitened using fair draws and the mean and median. The plot also shows the expected distribution of the variance for a reference $N(0,1)$ distribution. The data whitened using fair draws from the PSD posterior follows the expected distribution, while both the mean and median produce narrower vairaince distributions.

\bibliography{refs}

\end{document}